
\input harvmac

\def\LG{Lan\-dau-Ginz\-burg\ }

\def\half{{1 \over 2}}

\def\cmp#1{{\it Commun.\ Math. \ Phys.} \ {\bf #1\/}}
\def\nup#1{{\it Nucl.\ Phys.} \ {\bf B#1\/}}
\def\plt#1{{\it Phys.\ Lett.}\ {\bf #1\/}}
\def\ijmp#1{{\it Int. \ J.\ Mod. \ Phys.}\ {\bf A#1\/}}

\def\mpl#1{{\it Mod.\ Phys.\ Lett.} \   {\bf A#1}\ }

\def\coeff#1#2{\relax{\textstyle {#1 \over #2}}\displaystyle} 
coeff
\def\inbar{\vrule height1.5ex width.4pt depth0pt}
\def\IC{\relax\,\hbox{$\inbar\kern-.3em{\rm C}$}}
\def\IR{\relax{\rm I\kern-.18em R}}
\font\sanse=cmss12
\def\ZZ{\relax{\hbox{\sanse Z\kern-.42em Z}}}

\noblackbox

%
\def\Titletwo#1#2#3#4{\nopagenumbers\abstractfont\hsize=\hstitle
\rightline{#1}\rightline{#2}
\vskip .7in\centerline{\titlefont #3}
\vskip .1in \centerline{\abstractfont {\titlefont #4}}
\abstractfont\vskip .5in\pageno=0}

\def\Date#1{\leftline{#1}\tenpoint\supereject\global\hsize=\hsbody%
\footline={\hss\tenrm\folio\hss}}
%


\Titletwo{}{}
{Refining the Elliptic Genus$^*$ { \abstractfont
\footnote{}{$^*$ Work supported in part by funds provided by the DOE
under grant No. DE-FG03-84ER40168.}} } {}
\centerline{D. Nemeschansky \ and \ N.P. Warner}
\bigskip \centerline{Physics Department}
\centerline{University of Southern California}
\centerline{University Park}
\centerline{Los Angeles, CA 90089-0484.}
\vskip 1.0cm
We show how special forms of an $N=2$ Landau-Ginzburg potential
directly imply the presence of an $N=2$ super-$W$ algebra.
If the Landau-Ginzburg model has a super-$W$ algebra, we
show how the elliptic genus can be refined so as to give
much more complete information about the structure of the model.
We study the super-$W_3$ model in some detail, and present some
results and conjectures about more general models.
\vskip .1in
\vfill
\leftline{USC-94/002}
\leftline{hep-th/9403047}
\Date{March, 1994}
%


\newsec{Introduction}

\nref\WitLGa{E.~Witten, \nup{403} (1993) 159 .}
\nref \WitLGb{E.~Witten,``On the Landau-Ginzburg description of $N=2$
minimal
models,'' IASSNS-HEP-93-10, hep-th/9304026.}

\nref\LGboth{E.~Martinec, \plt{217B} (1989) 431; C.~Vafa and
N.P.~Warner,
\plt{218B} (1989) 51.}
\nref\EMart{E.~Martinec,  {\it ``Criticality, catastrophes and
compactifications,''}  in the V.G.~Knizhnik memorial volume, L.~Brink
{\it  et al.} (editors): {\it ``Physics and mathematics of
strings.''}.}
\nref\LVW{W.~Lerche, C.~Vafa and N.P.~Warner, \nup{324} (1989) 427.}
\nref\Gepner{D.~Gepner, \plt{222B} (1989) 207.}
\nref\HW{P.~Howe and P.~West, \plt{223B} (1989) 377.}
\nref\CGPA{S.~Cecotti,  L.~Girardello and A.~Pasquinucci, \nup{328}
(1989) 701.}

\nref\PDFSY{P.~di Francesco and S.~Yankielowicz, ``Ramond Sector
Characters
and $N=2$ Landau-Ginzburg Models,''SACLAY-SPhT 93/049,
hep-th/9305037;
P.~di Francesco, O.~Aharony and S.~Yankielowicz,
``Elliptic Genera and the Landau-Ginzburg approach to $N=2$
orbifolds,''
SACLAY-SPHT-93-068, hep-th/9306157.}
\nref\TKYYSKY{ T.~Kawai, Y.~Yamada and S.-K.~Yang, ``Elliptic
genera and $N=2$ superconformal field theory,'' KEK-TH-362,
hep-th/9306096.}
\nref\MHenn{ M.~Henningson, ``$N=2$ gauged WZW models and the
elliptic
genus,'' IASSNS-HEP-93-39, hep-th/9307040; P.~Berglund and
M.~Henningson,
``\LG orbifolds, mirror symmetry and the elliptic
genus,'' IASSNS-HEP-93-92.}
\nref\TKKM{T.~Kawai and K.~Mohri ``Geometry of (0,2) Landau-Ginzburg
Orbifolds,'' preprint KEK-TH 389, hep-th/9402148.}

Some important new results were obtained in \refs{\WitLGa,\WitLGb}
about the relationship between $N=2$ superconformal models and
$N=2$ Landau-Ginzburg  models \refs{\LGboth{--}\CGPA}.
In \WitLGb\ it was shown  that one could  compute the elliptic genus
of
the model by taking a
free field limit in which the coefficient of the potential vanishes.
{}From the elliptic genus one can then extract information about
Ramond
characters
of the  model and its orbifolds \refs{\PDFSY{--}\TKKM}.
A second result  of \WitLGb\ was to show that one could
obtain representatives of the superconformal
energy-momentum tensor in the Landau-Ginzburg model.
That is, even though Landau-Ginzburg action is not superconformal,
one is able to identify operators that, by virtue of the
Landau-Ginzburg equation of motion, generate the $N=2$
superconformal algebra upon left-moving states up to right-moving
states that are cohomologically trivial.
The idea being that these operators would generate the exact  $N=2$
superconformal algebra  at the infra-red fixed point of the model.

There are many natural extensions of this work, see for example,
\refs{\PDFSY{--}\TKKM}.
In this letter we wish to address the following issues.
First, the $N=2$ superconformal coset models based upon $CP_n$ have
very
particular Landau-Ginzburg potentials, and they also possess
super-$W$
algebras.
We will use the techniques of \WitLGb\ to establish a more direct
relationship between
the special form of the potential and the presence of such an
algebra.
Indeed, the classical limit of this  question has already been
extensively
analysed in
\ref\KM{K.~Mohri, ``N=2 Super-$W$ algebra in half-twisted
Landau-Ginzburg
model'', UTHEP-260, hep-th/9307029.}.
Our approach is a little different, and we will establish results for
the
quantum theory.

Secondly, given that there is such an extended algebra, one can, in
principle,
refine the elliptic genus so as to give information about the quantum
numbers of the $W$-charges in the Ramond sector.
We will indeed show how to modify the simple formulas for the
elliptic genus
given in \refs{\WitLGb{--}\MHenn} so as to extract detailed
information about
the $W$-structure of the Ramond ground states.

\newsec{Landau-Ginzburg formulation N=2 super-W algebras}

\nref\BW{J.~Wess and J.~Bagger, Supersymmetry
and Supergravity, Princeton University Press (second edition, 1992)}

Consider an $N=2$ supersymmetric Landau-Ginzburg model with action
\eqn\susyact{ S \ = \ \int \ d^2  x \ d^4 \theta \ \sum_j  \bar\Phi_j
\Phi_j -
\int \  d^2 x \ d^2 \theta\  W(\Phi_j) \ - \ \int \ d^2 x \ d^2 \bar
\theta \
W(\bar  \Phi_j) \ , }
where $\Phi_j$,  $j=1,\ldots , n$ are $N=2$ chiral superfields.
We will adopt the notation of \refs{\WitLGa,\WitLGb,\BW} and in
particular, super-derivatives are defined by:
\eqn\susyder{\eqalign{ D_+ \ &= \ {\del \over \del   \theta^+ } -i
\bar \theta^+ (\del_0+ \del_1) \ \ \ \ \  D_- \ = \ {\del \over \del
\theta^- } - i \bar \theta^- (\del_0 -\del_1) \ \ ;   \cr
\bar D_+ \ &= \ -{\del \over \del  \bar \theta^+ }
+ i\theta^+ ({\del_0+\del_1}) \ \ \ \ \ \bar D_- \ = \ -{\del \over
\del
\bar\theta^- } + i \theta^- ({\del_0-\del_1} )  \ . \cr}}
These super-derivatives satisfy the relations:
$\{ D_+ , \bar D_+ \}  =  2i ( \del_0 + \del_1)$, and
$\{ D_- , \bar D_- \} = 2i \ ( \del_0-\del_1) $.
Imposing chirality on the fields $\Phi_j$ means requiring that
$\bar D_+ \Phi_j = \bar D_- \Phi_j = 0$,
which implies that these superfields have an expansion:
\eqn\susyexp{\Phi_j(y, \theta) \ = \  \phi_j(y) \ + \  2 ~
\theta^{\alpha}\psi_{j, \alpha}(y)\ + \  \theta^{\alpha}
\theta_{\alpha} F_j
(y) \ ,}
where $\alpha \ = \ \pm\ $ and $y^m=  x^m + i\theta^\alpha
\sigma^m_{ \alpha,\dot \alpha}\bar \theta^{\dot \alpha} $.  Note that
we have normalized $\psi_{j,\alpha}$ differently from \WitLGb.

Given the kinetic term in \susyact, the short distance expansion of
$\Phi_j$ with $\bar \Phi_j$ is given by:
\eqn\phipro{ \Phi_j(x_1, \theta_1, \bar \theta_1) \bar
\Phi_j(x_2, \theta_2, \bar \theta_2) \ \sim \
-~ln ( \tilde x^m \tilde x_m) \ , }
where
\eqn\tilx{ \tilde x^m \ =\ ( x_1 -  x_2 )^m + i \theta_1 \sigma^m\bar
\theta_1+
i \theta_2 \sigma^m\bar \theta_2- 2 i \theta_1 \sigma^m\bar \theta_2
\ . }
One should note that in terms of the component fields, the foregoing
conventions lead to: $\phi_j(x) \ \bar \phi_j(0) \sim
-~ln ( x^m  x_m) $, and the rather non-standard form:  $\psi_{j,-}(x)
\ \bar \psi_{j,-} (0) \sim -i {1 \over (x^0 -x^1)} $.

The equations of motion derived from \susyact\  have a
very simple form:
\eqn\eqnmot{\eqalign{\bar D_+ \bar D_- \bar \Phi_j \ &= \ \half {
\del W\over
\del \Phi_i} \cr D_+ D_-  \Phi_j \ &= \ \half { \del W\over \del
\bar \Phi_i} \ .} }

Throughout this letter we will assume that the Landau-Ginzburg
potential is
quasi-homogeneous with indices $\omega_j$. That is,
\eqn\quasiho{W( \Phi_j) \ = \ \lambda^{-1}
W(\lambda^{\omega_j}\Phi_j)\ . }

The energy-momentum tensor, $T$, the supersymmetry generators,
$G^\pm$, and
the $U(1)$ current, $J(z)$, of a superconformal algebra can be
incorporated
into the various components of an $N=2$ superfield $\cal J$.
In \LG model one can explicitly construct a representative of the
superfield $\cal J$, whose components generate the the $N=2$
superconformal
algebra on the left-movers (up to trivial cohomology on the
right-movers) \WitLGb.  The superfield, ${\cal J}$, is simply:
\eqn\freej{ {\cal J} \ = \ \sum_j  \Big[\ \coeff{1}{2} (1-\omega_j)
D_-\Phi_j \bar D_-\bar \Phi_j -
i \ \omega_j \Phi_j (\del_0-\del_1) \bar \Phi_j \  \Big] \ \ , }
and it has been constructed so as to satisfy
\eqn\currentc{\bar D_+{\cal J} \ = \ 0 \ \ . }
This equation basically requires that ${\cal J}$ be holomorphic ( up
to the
cohomology of $\bar D_+$ ).
In particular, one has
\eqn\opejj{\eqalign{& {\cal J}(x_1,\theta_1, \bar \theta_1) \
{\cal J}(x_2,\theta_2, \bar \theta_2) \ = \cr  &
-{c \over  3 \tilde x_{12}^2} ~+~ 2~\Bigl (
{\theta^-_{12}\bar \theta^-_{12} \over \tilde x^2_{12}}
+ {i \over 2} { \theta_{12}^- \over \tilde x_{12}}  D_-
+ {i \over 2} { \bar\theta_{12}^- \over \tilde x_{12}}
\bar D_-+ {\theta_{12}^- \bar\theta_{12}^- \over \tilde x_{12} }
(\del_0
-\del_1) \Bigl ){\cal J}(x_2,\theta_2) \ ,\cr }}
where
$\tilde x_{12} \ = \ z_1-z_2 + i(\bar\theta_1^-\theta_2^-+
\theta_1^-\bar\theta^-_2)$ , $ \theta_{12}^-\ = \
\theta_1^--\theta_2^-$ and
$z=x^0-x^1$  and  $c$ is the  charge of the $N=2$ supersymmetric
model.
The normalization of ${\cal J}$ has been determined by fixing the
leading
singularity in \opejj.

Note that in order to establish  \currentc\ one needs to use
quasi-homogeniety
of $W$ along with the equations of motion \eqnmot.

\nref\KS{Y.~Kazama and H.~Suzuki, \nup{234} (1989) 73.}
\nref\Fusion{D.~Gepner, \cmp{141} (1991) 381; K~Intriligator, \mpl{6}
(1991) 3543.}

For $n=1$ and $W(\Phi) \ = \ {1 \over k+2} \Phi^{k+2}$, this
Landau-Ginzburg
model describes the $N=2$ superconformal minimal  models with central
charge
$c=3k/(k+2)$ \refs{\WitLGb,\LGboth}.
The currents in the superfield $\cal J$  constitute a complete chiral
algebra
for the theory, that is, the Hilbert space is finitely reducible as
representation of the algebra.
When $n \ge 2$, this is no longer true.
However, for special Landau-Ginzburg potentials we know that the
chiral
algebra can be extended to an $N=2$ super-$W$ algebra.
In particular the  $N=2$  superconformal coset models \KS:
\eqn\KSmodel{ {SU_k(n+1) \times SO_1(2n) \over SU_{k+1}(n) \times
U(1) }}
have an $N=2$ super-$W$ algebra, and this generally believed to be a
complete
chiral algebra.
Moreover, these models have a Landau-Ginzburg formulation.
An easy way to compute the Landau-Ginzburg potential is as  follows
\refs{\LVW,\Fusion}.
The potential, $W$, is given by
\eqn\defiwpot{W \ = \  \sum_{p=1}^n {1 \over k+n+1} ~\xi_p^{k+n+1}}
where the $\Phi_j$ are defined by
\eqn\defxip{\Phi_j\  =\
\sum_{1 \le  p_1 <p_2<\ldots p_j \le n} \xi_{p_1}\xi_{p_2}
\ldots \xi_{p_j} \ .}
Since $W$ is a symmetric function of the $\xi_p$,
one can write $W$ as a function of
$\Phi_j$ and then $W(\Phi_j)$ is the requisite Landau-Ginzburg
potential.

We note that $W$ is uniquely characterized (up to scaling of the
$\Phi_j$)
by its quasi-homegeneity and the differential equation
\eqn\diffeq{{\del^2 W \over \del \xi_p \del \xi_q}\ = \ 0 \ \ \ \
p\ne q\ . }
This implies obvious second order differential equations in terms of
the fields
$\Phi_j$.
In particular, for $n=2$, the Landau-Ginzburg potential is uniquely
characterized by the scaling indices $\omega_1 \ = {1\over k+3} $ and
$\omega_2 \ = {2\over k+3} $ and the differential equation,
\eqn\diff{ {\del^2 W \over( \del \Phi_1)^2 } \ + \  \Phi_1 {\del^2 W
\over\del
\Phi_1 \del \Phi_2 } \ + \  \Phi_2 {\del^2 W \over( \del \Phi_2)^2 }
\ + \
{\del W \over \del \Phi_2 } \ = \ 0 \ .}

For simplicity we will restrict our discussion to super-$W_3$
generators,
but the
generalization to higher spin elements of the chiral algebra should
be
straightforward though algebraically awful\foot{Note that we have not
yet restricted the number of superfields, we have simply focussed
upon
the simplest non-trivial extension of the chiral algebra.}.
One can determine the super-${\cal W}_3$ current
in much the same way as one determines
the current $\cal J$.  One can make an Ansatz as follows:
The lowest component, $S$, of the superfield ${\cal W}_ 3$
has dimension two and
therefore the realization of ${\cal W}_3$ in terms  of the chiral
superfields
must consist of terms with four super-derivatives.  As was the case
for the
current $\cal J$, the fields $\bar \Phi_i, i=1,2$, never appear
without a
super-derivative.  Futhermore, the number of the chiral fields,
$\Phi_i$,
is equal
to the number  of anti-chiral fields, $\bar \Phi_i$.
These   constraints leave one  with about twenty possible terms.
The constraint \currentc\ was solved in the classical limit in \KM\
for a
number
of $W_n$ generators, (and not just $W_3$).
Here we will show how to greatly simplify the Ansatz, and arrive at
the full
quantum result for $W_3$.
{}From this we will be able to conjecture the result for general
$W_n$.

Recall that the models \KSmodel\  factorize into a tensor products
according to
\eqn\decom{ {\cal M}_1 \ \times\  {\cal M}_2 \ \times \ {\cal M}_3 \
= \
{SU_{k}(n+1)\over SU_k(n)\times U(1)} \ \times {SU_k(n) \times
SU_1(n)\over
SU_{k+1}(n) } \ \times \ U(1) \ .}
Let $T_1$ and $T_2$ denote the energy-momentum tensors of ${\cal
M}_1$ and $
{\cal M}_2$,
respectively.
The corresponding central charges are $ c_1 \ = \
{ n(k-1)(1+2k+n)\over (k+n)(k+n+1)}$ and $ c_2 \ = \  (n-1)(1-{
n(n+1)\over
(k+n)(k+n+1)}) $.
The lowest component, $S$, of the $N=2$ super-$W_3$ generator can the
be
written, up to normalization, as
\eqn\sint{ S \ = \ c_2 T_1 - c_1 T_2 \ . }
The field $S$ manifestly  has vanishing operator product with $J$, as
is
required by the super-$W_3$ algebra.
The relative coefficient of $T_1$ and $T_2$ in \sint\ is determined
by
requiring that $S$ be a good conformal field, ({\it i.e.} with no
anomalies).
Our aim will be to construct a superfield $ {\cal T}_2$, with lowest
component
$T_2$. Once we have $T_2$, we can reconstruct
$S$ by writing,
\eqn\news{S \ = \  c_2 T\ - {3 c_2 \over 2 c} \ J^2\ -(c_1+c_2) \ T_2
\ , }
where $T$ is the energy-momentum tensor of the complete model.
This then implies that the ${\cal W}_3$ generator is given by
\eqn\supewthree{{\cal W}_3 \ = \ -{ic_2 \over 4 }(D_-\bar  D_- -\bar
D_-D_-)
{\cal J} \ +  \ { 3c_2 \over 2c }  {\cal J}^2 \
-\ (c_1+c_2){\cal T}_2 \ . }

We now specialize to $n=2$, for which ${\cal M}_2$ is a standard
minimal model.
In making an Ansatz for ${\cal T}_2$ it is natural to assume that the
$N=2$
superfields provide the standard realization of the minimal model in
terms of a
single free boson.
We, therefore, make an Ansatz for the superfield, $\hat{\cal J},$
corresponding
to this free boson.  Apart from the fact it works, we have another
reason for making this Ansatz and we will comment about this later.
The most general form for $\hat{\cal J}$, consistent with rules
outlined
above, is:
\eqn\hatj{\hat{\cal J} \ = \ a D_-\Phi_1 D_-\bar \Phi_1 +
 b  D_-\Phi_2\bar D_-\bar \Phi_2 + c \Phi_1 \del \bar \Phi_1 +
d\Phi_2 \del \bar \Phi_2 \ ,}
where $\partial = \partial_0 - \partial_1$.
The coefficients $a,b,c$ and $d$ in \hatj\ are determined by
requiring that
$\hat{\cal J}$ have proper operator product expansion with $T$, and
imposing
the operator equation of motion
\eqn\opeeq{\bar D_+ {\cal T}_2 \ = \ 0 \ . }
We find
\eqn\hatjfi{\hat{\cal J} \ = \ \ {i \over 2} \sqrt{1 - \omega}~
D_-\Phi_1
D_-\bar \Phi_1 ~-~ {i \over 2}{ 1 \over \sqrt{1 - \omega}}~
D_-\Phi_2\bar
D_-\bar \Phi_2 ~+~ { \omega \over \sqrt{1 - \omega}}~ \Phi_1
\del \bar \Phi_1  \ ,}
where $ \omega \ = \ {1 \over  k+3}$.
For future reference we note  that lowest component of the $U(1)$
current is
\eqn\uonecur{ \hat j \ = \ {1 \over \sqrt{1-\omega}} (\omega \phi_1
\del \bar \phi_1 \ - \ i (1 - \omega) \bar \psi_{1,-}\psi_{1,-} \ + \
i \bar \psi_{2,-} \psi_{2,-}) \ . }
It is important to note that  we could not impose the operator
equation of
motion
on $\hat {\cal J}$ itself because $\hat j(z)$ does not commute with
the
screening charges and is not an operator in the minimal model.

\nref\DNSY{D.~Nemeschansky and S.~Yankielowicz,``$N=2$ W-algebras,
Kazama-Suzuki Models and Drinfeld-Sokolov Reduction'', USC-preprint
USC-007-91 (1991)}
\nref\KIto{K.~Ito, \plt{259} (1991) 73.}

There are several important aspects to the foregoing computation.
First, the equation \opeeq\ is only satisfied by the virtue of
quasi-homogeneity
\quasiho, the Landau-Ginzburg equations \eqnmot\ and the constraint
\diff\
upon the Landau-Ginzburg potential.
In particular, if we had not used the constraint \diff, then there
would be no
general solution.
The second point is that the verification that ${\cal T}_2$ satisfies
\opeeq\
is complicated by the operator ordering after one uses  the
Landau-Ginzburg
equation of motion, and some subtleties of screening operators in the
minimal
model.
In this computation, it  is elementary to fix  $a,b,c$ and $d$, but
to verify
that ${\cal T }_2$ satisfies the operator equation \opeeq\ to all
orders in the
Wick contractions is much more complicated.
Indeed we ultimately confirmed our results by finding the translation
table
between the Landau-Ginzburg fields of \WitLGb\ and the explicit
formula
for ${\cal W}_3$
given in terms of Drinfeld-Sokolov reduction in  \refs{\DNSY,\KIto}.
These issues will be discussed fully in \ref\NW{D.~Nemeschansky and
N.P.~Warner, to appear in the near future.}.

\newsec{The elliptic genus}

\nref\SWEG{A.N.~Schellekens and N.P.~Warner,
\plt{177B} (1986) 317; \plt{181B} (1986) 339; \nup{287}  (1987)
317;  K.~Pilch, A.N.~Schell\-ekens and N.P.~Warner,\nup{287} (1987)
362.}
\nref\WitEG{E.~Witten, \cmp{109} (1987) 525; ``The index of the Dirac
Operator in Loop Space,'' in P.~Landweber, ed.,
{\it Elliptic Curves and Modular Forms in Algebraic Topology,},
Lecture
Notes in Mathematics 1326,  (Springer-Verlag, 1988)}
\nref\AKMW{O.~Alvarez, T.-P. Killingback, M.~Mangano, and
P.~Windey, {\it Nucl.\ Phys. B (Proc. Suppl.)} \ {\bf 1a\/} (1987)
189;  \cmp{111} (1987).}
\nref\mathellgen{S.~Ochanine, ``Genres elliptiques \'equivariants,''
in P.~Landweber, ed., {\it Elliptic Curves and Modular Forms in
Algebraic Topology,} Lecture
Notes in Mathematics 1326,  (Springer-Verlag, 1988);
P.~S. Landweber, ``Elliptic cohomology and modular forms,'' and
``Elliptic genera: An introductory overview,'' {\it ibid.};
D.~Zagier, ``Note on the Landweber--Stong elliptic genus,''
{\it ibid.} .}

The elliptic genus of the  model \susyact\ is defined by
\refs{\SWEG {--} \mathellgen}:
\eqn\ellgen{E(q,\gamma) ~=~ Tr_{\cal H} ~\left( (-1)^F  q^{H_L}~ \bar
q^{H_R}
{}~ exp(i \gamma J_{0}) \right) \ .}
In this expression $\cal H$ is the complete Hilbert space of the
model in
the Ramond sector, $H_L$ and $H_R$ are the hamiltonians of the
left-movers
and right-movers, $F$ is the total fermion number, and $J_0$ is the
left-moving  $U(1)$ charge.  Contrary to the conventions of \WitLGb,
we will
identify $H_L$ with the Virasoro generator $L_0$, and $H_R$ with
$\bar L_0$.
The standard index argument can be used to show that in the
right-moving
sector,
only the ground-states contribute to the trace.  As a result, the
elliptic
genus is a function of $q$ alone (and not a function of $\bar q$),
and consists
of a sum of the (left-moving) Ramond ground-state characters.

The $N=2$, $U(1)$ current is given obtained from the lowest component
of \freej.  Specifically, one has
$J = -i {\cal J} |_{\theta = \bar \theta =0}$, and so:
\eqn\Jform{J(z) ~=~ \sum_{j=1}^n ~\left[~ i(1 - \omega_j)~\bar
\psi_{j,-}(z)
{}~\psi_{j,-}(z) ~-~ \omega_j~\phi_j(z) ~\partial \bar \phi_j(z)
{}~\right] \ .}
The action of the charge $J_0$ on the (left-moving) superfield
components is:
\eqn\Jaction{\eqalign{\phi_j ~&\to~ exp(i\omega_j \gamma)~ \phi_j \cr
\psi_{j,-} {}~&\to~ exp(i(\omega_j - 1) \gamma) ~ \psi_{j,-} \ ,}}
where $\gamma$ is the parameter.

One of the key observations in \WitLGb\ was that the elliptic genus
could
be computed in a \LG model by taking the limit in which the
coefficient of
the \LG potential goes to zero, and the \LG fields become free.  One
therefore obtains a simple free-field expression for $E(q,\gamma)$:
\eqn\ellform{E(q,\gamma) ~=~ e^{-i \gamma {c \over 6}} ~\prod_{j=1}^n
{}~ \prod_{p=1}^\infty {{(1 - q^{p-1} e^{i \gamma (1 - \omega_j)})
(1 - q^p e^{- i \gamma (1 - \omega_j)})} \over {(1 - q^{p-1} e^{i
\gamma
\omega_j}) (1 - q^p e^{-i \gamma \omega_j})}}  \ .}
In this expression $c$ is the central charge of the model, and is
given by
\refs{\LGboth,\LVW,\HW}:
\eqn\centch{ c ~=~ 3~\sum_{j=1}^n ~(1 - 2 \omega_j) \ .}
It was verified in
\refs{\PDFSY{--}\MHenn} that for $\omega_j = j/(k+n+1)$, this is
indeed
the sum of the Ramond ground state characters of the $N=2$
superconformal
coset models \KSmodel.

The elliptic genus \ellgen\  has already been refined in the sense
that  it gives the $N=2$, $U(1)$ charges of the states.
For the minimal models
($n=1$), knowledge of the $U(1)$ charges is sufficient to isolate
individual
Ramond characters from the elliptic genus \WitLGb.  This is no longer
true
when one has more superfields ($n > 1$).   However, if such a model
posseses
an $N=2$ super-$W$ algebra one should be able to once again resolve
the
elliptic genus into individual characters.
To accomplish this, one seeks left-moving
generators of the extended chiral algebra that commute with each
other and
with $J_{0}$ and $H_L$.  One can then, in principle, insert
exponentials
of these additional charges into the elliptic genus and completely
refine it
with respect to the extended algebra\foot{It is, of course, critical
that
one only insert into the elliptic genus operators that commute with
the
right-moving supercharge, otherwise the elliptic genus would no
longer be
an index, and would depend upon $\bar q$.  This is why we have
restricted
to the left moving chiral algebra here, but we note that there are
discrete
exponentials of right-moving charges that also commute with the
right-moving
supercharges.}.  The obvious problem now is to find the
generalization of
\ellform.  We will illustrate the procedure by restricting our
attention to
two superfields and the $N=2$ super-$W_3$ algebra.

\nref\LNW{W.~Lerche, D.~Nemeschansky and N.P.~Warner, unpublished.}

Let $S_0$ be the zero-mode of the left-moving spin-2 field
that is the lowest component of the $W_3$-supermultiplet.  This
charge
commutes with $J_0$ and $H_L = L_0$, and so one can define an obvious
refinement by inserting $p^{S_0}$ into \ellgen.  It is also
relatively
easy to see that the quantum numbers of $S_0$ and $J_0$ are
sufficient to
resolve the Ramond ground-states in the $N=2$ super-$W_3$ model \LNW.

There does not appear to be a direct way to find a simple expression
for
this refinement of the elliptic genus.  Instead we will construct a
simpler
character function with equivalent information.  This approach will
lead us to a simple generalization of the elliptic genus, and to a
method
that will easily generalize to higher super-$W$ algebras.
The first step is to use the fact that the $N=2$ superconformal model
factorizes as in \decom.  We then refine the elliptic genus using the
operator
$L_0^{(2)}$, which is the zero-mode
of the energy-momentum tensor, $T_2$, of ${\cal M}_2$ in \decom.
That is,
we define
\eqn\refell{E(q,p,\gamma) ~=~ Tr_{\cal H} ~\left( (-1)^F  q^{H_L}~
\bar
q^{H_R} ~p^{L_0^{(2)}} ~ exp(i \gamma J_{0})  \right) \ .}
As we saw in the previous section, the energy momentum tensor, $T_2$,
appears in the superconformal model in terms of its standard
realization
in terms of a single boson, with associated $U(1)$ current $\hat j_0$
defined by \uonecur.  Introduce the function:
\eqn\ellchar{F(q,\nu,\gamma) ~=~ Tr_{\cal H} ~\left( (-1)^F  q^{H_L}~
\bar
q^{H_R} ~ exp(i \nu \hat j_{0}) ~ exp(i \gamma J_{0})  \right) \ ,}
and define its symmetrized form by:
\eqn\Fsymm{F_s (q,\nu,\gamma) ~=~ F(q,\nu,\gamma) ~+~ F(q,\nu,
-\gamma) \ .}
It is this function (and its generalizations) that is easily computed
in the free field limit of the \LG model.  The problem is that,
unlike $T_2(z)$, the $U(1)$ current
$\hat j(z)$ is emphatically not in the chiral algebra of the $N=2$
superconformal model.  The current $\hat j(z)$ does not satisfy any
operator equations analogous to \currentc, or equivalently it does
not commute
with the requisite screening currents.
Therefore $F(q,\nu,\gamma)$ is not going to be any kind of character
on the
Hilbert space of the $N=2$ superconformal model.  However, the
function
$F_s$ is a kind of character on the $N=2$ superconformal
model, and it contains exactly the same information as
$E(q,p,\gamma)$.
Indeed, one has
\eqn\inttrans{E(q,p,\gamma) ~=~ \sqrt{2 \over \pi} ~ {\eta(q) \over
\eta(pq)} ~ \int_{- \infty}^\infty ~ e^{-\nu^2 \lambda}~
F_s(q,\gamma,
\nu - \coeff{a}{\lambda}) ~d\nu \ .}
The gaussian integral has the effect of replacing $exp(i \nu \hat
j_{0})$
by $p^{\half (\hat  j_{0} + a)^2 - \half a^2}$, where
$p=e^{-1/\lambda}$.
Thus for  the proper choice of $a$, each state in the trace is
weighted by
the power of $p$ appropriate to the energy of the associated
$\hat j_0$-momentum
state.  The $\eta$-function pre-factors in \inttrans\ take care of
the
oscillator contribution to the minimal model. One can invert
the integral transform \inttrans\ by essentially performing the
inverse
Laplace transform.  The symmetrization of $F$ is necessary because
$E(q,p,\gamma)$  is an even function of the $\hat j_0$ eigenvalues,
and so
the inversion of \inttrans\ must yield an even function of $\nu$.

\nref\BKos{B.~Kostant,  {\it  Am.\ J.\ Math.} \ 81 (1959) 973.}
\nref\VFSL{V.A.~Fateev and S.L.~Lukyanov, \ijmp{3} (1988) 507;
``Additional symmetry and exactly-soluble models in two
dimensional conformal field theory'', Landau Institute preprint
 (1988).}

To understand more generally what is transpiring here, we recall a
basic
theorem about Lie algebras \BKos:
Two weights of a Lie alegebra are equal
{\it up to Weyl rotations} if and only if all the Casimir invariants
take the same values on the two weights.
Essentially, if we have a bosonic
realization of a $W$-algebra then the values of the $W$-charges on
the
bosonic momentum states are precisely the values of the Casimirs of
the
underlying Lie Algebra \VFSL.  Thus knowing the $W$-charges of the
momentum
states is equivalent to knowing the {\it Weyl symmetrized} character
of the
bosonic Hilbert space.

Therefore, because the function $F_s$ is precisely equivalent to a
refined form of the elliptic genus, one can also compute it for the
$N=2$ superconformal model by taking the free field limit in which
the coefficient of the \LG potential vanishes.

Let  $y=exp[{i {{\omega \nu} \over \sqrt{1 -\omega}}}]$ and
$z=exp[{i \omega \gamma}]$, then, in the free field limit one has:
\eqn\Ffree{\eqalign{F(q,y,z) ~=~  y^{-1} z^k ~ \prod_{p=1}^\infty ~
\Bigg\{ ~ & {{(1 - q^{p-1} y^{-(k+2)} z^{(k+2)})~(1 -
q^p y^{(k+2)} z^{-(k+2)}) } \over {(1 - q^{p-1} y^{-1} z) ~
(1 - q^p y z^{-1}) } }
\cr & {{(1 - q^{p-1} y^{(k+3)} z^{(k+1)}) ~(1 - q^p y^{-(k+3)}
z^{-(k+1)}) }
\over  {(1 - q^{p-1} z^2) ~(1 - q^p z^{-2}) } } ~ \Bigg\}  \ . }}
One can then extract the refined elliptic genus from:
\eqn\refell{F_s(q,y,z) ~=~ F(q,y,z) ~+~ F(q,y^{-1},z) ~=~ F(q,y,z)
{}~+~ F(q,y,z^{-1}) \ .}

Alternatively, and perhaps more usefully, one can use \refell\ to
generate  the characters of the first factor, ${\cal M}_1$, in
\decom.
That is, if one expands $F_s(q,y,z)$ and collects the coefficient of
$y^a z^b$, then this will  be $(\eta(q))^2$ times the character of a
representation, $R$, of the model ${{SU_k(3)} \over {SU_k(2) \times
U(1)}}$.  This representation, $R$, is the one that is paired in the
$N=2$ superconformal model with a state that has $N=2$, $U(1)$ charge
$b \over {k+3}$ and minimal model momentum
$a \over \sqrt{2(k+2)(k+3)}$.
We have used {\it Mathematica} to verify this explicitly for
$k=1,2$ and powers up to $q^4 y^{10} z^{10}$.

Thus one can extract the characters of the component models directly
and easily from the elliptic genus.

\nref\DGep{ D.~Gepner,
``A Comment on the Chiral Algebra of Quotient Superconformal Field
Theory'', Princeton preprint PUPT 1130 (1989); \cmp{142} (1991) 433.
}

One can also play various other games.  For example, one can extract
the complete $N=2$ super-$W_3$ character above a {\it single} Ramond
ground state.  Recall that the Ramond ground states of \KSmodel\ are
in
one-to-one correspondence with the $SU_k(3)$ highest weight labels
\refs{\LVW,\DGep}.  One can obtain a Ramond ground state from
a highest weight state, $\Lambda$, of $SU_k(3)$ by tensoring it with
the $SO(4)$ spinor ground state of maximum charge, and then factoring
out the $SU_{k+1}(2) \times U(1)$ highest weight state that is simply
the projection onto $SU(2) \times U(1)$ of the sum of the $SU(3)$ and
$SO(4)$ weights.  Thus, if $\Lambda$ has Dynkin labels $(n_1,n_2)$,
then the $SU(2)$ Dynkin label is $n_1$, and the $U(1)$ charge  is
$n_1 + n_2 + 3$ (the shift of $+3$ comes from the $SO(4)$ spinor
contribution).  Therefore, since the elliptic genus consists of
purely Ramond ground state characters, we may isolate a particular
such ground state by fixing the representation of
$SU_{k+1}(2) \times U(1)$.  Observe that in the decomposition \decom,
the factor $SU_{k+1}(2)$ appears in the denominator of the
minimal model.  Moreover, in the standard bosonic formulation, the
minimal model momentum is:
\eqn\minmodp{p ~=~ \sqrt{\coeff{k+2}{2(k+3)}} ~~m_2 ~-~
\sqrt{\coeff{k+3}{2(k+2)}} ~~m_1 ~+~ \sqrt{2(k+2)(k+3)} ~~ m \ ,}
with $m \in \ZZ$,	$m_1$ $= 1,2,\dots, (k+1)$ and 	$m_2$ $= 1,2,
\dots, (k+2)$.  Such a bosonic momentum state contributes to the
minimal model character of the $\Phi_{m_1,m_2}$ representation.
To fix the label of $SU_{k+1}(2)$ we need to fix $m_2$, but to obtain
the complete character, we must at the same time sum over all
allowed values of $m_1$.  Take
$\nu = 2 \pi i \sqrt{{{k+2} \over {k+3}}} j$ and observe that
with this choice one has $e^{i \nu \hat j_0} = e^{i \sqrt{2} p \nu}
= e^{- 2\pi ij m_2/(k+3)}$.
Summing over $j$ will then project the elliptic genus onto a
specific $SU_{k+1}(2)$ state.  Therefore, the function
\eqn\ellproj{E_\ell(q,\gamma) ~=~ {1 \over k+3} ~ \sum_{j=0}^{k+2}
{}~ e^{ 2\pi ij (\ell + 1)/(k+3)} ~ F_s \left( q,\gamma,
\nu = 2 \pi i j \sqrt{\coeff{k+2}{k+3} }  \right) \ }
provides the projection onto the Ramond ground states with $SU(2)$
Dynkin label equal to $\ell$.  The $U(1)$  quantum number can also
be fixed, exactly as was done in \WitLGb, by performing a similar
finite sum over values of the parameter $\gamma$.

\newsec{Generalizations}

It is relatively simple to conjecture about the generalization of
results to models of the from \KSmodel\ for $n\ge 3$.  The first step
is
to seek out the free bosonic realization of ${\cal M}_2$ within the
$N=2$ superfields.  For $\ell = 1, \dots, n-1$, define:
\eqn\conjhatj{ \hat j_\ell(z) ~=~ {1 \over \sqrt{1 - \omega}}~ \big[
{}~\omega ~ \phi_\ell \partial \bar \phi_\ell ~-~  i (1 -
\omega)~\bar
\psi_{\ell,-}(z) ~\psi_{\ell,-}(z) ~+~ i \bar \psi_{\ell+1,-}(z)
{}~\psi_{\ell+1,-}(z)~\big] \ ,}
where $\omega = 1/(k+n+1)$.
Observe that these currents are orthogonal to $J(z)$, and satisfy
$$ \hat j_\ell(z)~\hat j_m(w) ~=~ {A_{\ell m} \over (z-w)^2}
{}~+~ \dots \ ,$$
where $A_{\ell m}$ is the Cartan matrix of $SU(n)$\foot{Remember that
we have the somewhat unusual conventions: $\phi_j(x) \ \bar \phi_j(0)
\sim -~ln ( x^m  x_m) $,  $\psi_{j,-}(x) \ \bar \psi_{j,-} (0) \sim
-i {1 \over (x^0 -x^1)} $.}.  We may therefore
set $\hat j_\ell(z) = \vec \alpha_\ell \cdot \partial \vec X(z)$,
where the $\vec \alpha_\ell $ are the roots of $SU(n)$ and $\vec
X(z)$
is a vector of $n$ free bosons.  We have not yet
proved, but have a compelling body of evidence that these bosons
generate,
inside the $N=2$ Hilbert space, the standard free bosonic realization
of
the $W_n$ minimal model, ${\cal M}_2$ in \decom.  We also have
confirmation
of this from the corresponding refinements of the elliptic genus.

Define the character
\eqn\genellchar{F(q,\nu,\gamma) ~=~ Tr_{\cal H} ~\bigg( (-1)^F
q^{H_L}~
\bar q^{H_R} ~ exp\Big(i \sum_\ell \nu_\ell \hat j_{\ell,0}\Big) ~
exp(i \gamma J_{0})  \bigg) \ ,}
and symmetrize it with respect to the Weyl group of $SU(n)$:
\eqn\genFsymm{F_s (q,\nu,\gamma) ~=~ \sum_{w \in W(SU(n))}
F(q,w(\nu),\gamma)  .}
As above, we claim that this can be computed in the \LG model
in the limit where the coefficient of the potential vanishes.
Therefore, we have
\eqn\Fastheta{F(q,\nu,\gamma) ~=~  \prod_{j=1}^n ~
{\vartheta_1(a_j|\tau) \over \vartheta_1(b_j|\tau) } \ ,}
where:
\eqn\abdefn{\eqalign{a_j ~&=~ (1 - \omega) \nu_j ~-~ \nu_{j-1} ~+~
(1 -j \omega) \gamma \cr b_j ~&=~  - \omega \nu_j  ~-~
j \omega \gamma \ ,}}
and $\nu_0 \equiv \nu_n \equiv 0$.

Using {\it Mathematica}, we have checked the expansion of $F_s$
explicitly for $n=3,4; k=1,2$
and find that it does indeed produce the proper generalization of the
results of the previous section.  It is also amusing to note that for
$n \ge 3$ it is far from obvious that $F_s$ is non-singular as
$\gamma \to 0$.  For $n=3$ one can write the six terms in $F_s$ over
a common denominator, and the numerator becomes sums of products of
four theta functions.  The numerator vanishes in the limit $\gamma
\to 0$
by virtue of the vanishing of a particular sum of three terms each
consisting of a product of
four theta functions.  It is this same identity that is
of particular importance in establishing that elliptic Boltzmann
weights
satisfy the Yang-Baxter equations.

\newsec{Conclusion}

In this letter we have refined the elliptic genus for $N=2$
superconformal
models by including new charges arising from the super-$W$ algebra of
the
Landau-Ginzburg models.
We have argued that we could find these new generators only when the
superpotential has a very specific form  and  satisfies additional
second order differential equations such as \diff.

The refinement of the elliptic genus enabled us to isolate
characters of
various component parts of the $N=2$ superconformal Hilbert space,
and also
isolate individual characters in the Ramond sector.

\nref\FGMP{P.~Fr\'e, F.~Gliozzi, M.~R.~Monteiro and A.~Piras, {\it
Class.
Quantum. Grav.} {\bf 8} (1991) 1455;
P.~Fr\'e, L.~Girardello, A.~Lerda and P.~Soriani, \nup{387} (1992)
333.}

In writing this letter we have suppressed many of the technical
details.
A very useful guide in our computations has been the translation
table between
the Landau-Ginzburg fields of \refs{\WitLGa,\WitLGb} (which is
essentially the
same as $\beta, \gamma$-system of \refs{\FGMP,\TKYYSKY} ) and the
fields that
naturally appear in Drinfield-Sokolov reduction.
Questions about the operator equations of motion in the
Landau-Ginzburg
formulation can then be converted into fairly standard questions
about
commutations with screening charges.  It was also detailed
knowledge of the relationship between the \LG fields and the
superfields of Drinfled-Sokolov reduction that led us to make the
Ansatz for $\hat {\cal J}$ instead of working with a much more
complicated Ansatz for ${\cal W}_3$.

We have also suppressed a rather interesting technical point about
the
embedding of the bosonic formulation of minimal models into the $N=2$
superfield Hilbert space.
The minimal model screening charges are slightly non-standard, and
this directly linked with the fact the decomposition \decom\ is not
just a simple tensor product, but there is a ``locking together'' of
 representations of $ {\cal M}_1$ and $ {\cal M}_2$ so as to make a
non-trivial modular-invariant. All of the foregoing
issues will be discussed in detail \NW.

Our purpose in this letter has thus been to distill the essential
ideas and
some key results of our work and defer the technical details, and
some of the
subtleties, to a future publication \NW.

\bigskip

\leftline{\bf Note added:}
\medskip
While working on this manuscript, we were
advised that W.~Lerche and A.~Sevrin had also derived results
related to ours about the connection between super-$W$ algebras
and the form of the superpotential.

\vfill
\listrefs
\eject
\end